\documentclass{PoS}
\usepackage{psfig}

\title{SKA Key science project: Radio observations of cosmic
reionization and first light}

\ShortTitle{Cosmic reionization and first light}

\author{\speaker{C.L. Carilli}\thanks{The National Radio Astronomy
Observatory (NRAO) is operated by Associated Universities, Inc. under
a cooperative agreement with the National Science Foundation.}\\
        National Radio Astronomy Observatoty, Socorro, NM, USA\\
        E-mail: \email{ccarilli@nrao.edu}}

\abstract{I update the SKA key science program (KSP) on first light
and cosmic reionization. The KSP has two themes: (i) Using the 21cm
line of neutral hydrogen as the most direct probe into the evolution
of the neutral intergalactic medium during cosmic reionization. Such
HI 21cm studies are potentially the most important new window on
cosmology since the discovery of the CMB.  (ii) Observing the gas,
dust, star formation, and dynamics, of the first galaxies and AGN.
Observations at cm and mm wavelengths, provide an unobscured view of
galaxy formation within 1 Gyr of the Big Bang, and are an ideal
complement to the study of stars, ionized gas, and AGN done using
near-IR telescopes. I summarize HI 21cm signals, challenges, and
telescopes under construction. I also discuss the prospects for studying
the pre-galactic medium, prior to first light, using a low frequency
telescope on the Moon. I then review the current status of mm and cm
observations of the most known distant galaxies ($z \ge 6$). I make
the simple argument that even a 10\% SKA-high demonstrator will have a
profound impact on the study of the first galaxies. In particular, 
extending the SKA to the 'natural' atmospheric limit (set by the O$_2$
line) of $\sim 45$GHz, increases the effective sensitivity to
thermal emission by another factor four.}

\FullConference{From planets to dark energy: the modern radio universe\\
		 October 1-5 2007\\
		 University of Manchester, Manchester, UK}

\begin{document}

\section{Introduction}

Cosmic reionization corresponds to the transition from a fully neutral
intergalactic medium (IGM) to an (almost) fully ionized IGM caused by
the UV radiation from the first luminous objects.  Reionization is a
key benchmark in cosmic structure formation, indicating the formation
of the first luminous objects.  Reionization, and the preceding 'dark
ages', remain the last of the major phases of cosmic evolution left to
explore. 

The SKA international science advisory committee identified 
cosmic reionization and first light as one of the key science projects
for a future square kilometer array. The KSP had two parts:

\begin{itemize}

\item  Use the 21cm line of neutral hydrogen as the most direct
probe into the evolution of the neutral intergalactic medium during
cosmic reionization. Such HI 21cm studies are potentially the most
important new window on cosmology since the discovery of the CMB.

\item Observe the gas, dust, star formation, and dynamics, of the
first galaxies and AGN.  Observations at cm and mm wavelengths,
provide an unobscured view of galaxy formation within 1 Gyr of the Big
Bang, and are an ideal complement to the study of stars, ionized gas,
and AGN done using near-IR telescopes.

\end{itemize}

I should start with a mea culpa: the original SKA KSP was misnamed as
study of the 'dark ages'.  The dark ages correspond to the period
prior to reionization and first light, while the SKA KSP focuses on
essentially cosmic reionization, which signals the end of the dark
ages. This review will focus again on reionization, although I have
included a summary of recent work suggesting that the dark ages
themselves may be a gold mine of cosmological discoveries through the
HI 21cm studies of the pre-galactic medium, potentially studied with a
low frequency radio telescope on the Moon (section 3.5).

Since this KSP was first proposed (Carilli et al. 2004), there has
been an explosion in theoretical studies of the expected HI 21cm
signal from reionization, as well as extensive
reviews written on the first galaxies and the process of reionization
(Furlanetto, Oh, Briggs 2007; Ellis 2007; Fan, Carilli, Keating 2007).
In this review, I will concentrate on updating predicted signals,
observational challenges, and telescopes under construction. More on
the theory of reionization can be found in Ferrara (this volume).

In Section 4, I present the latest results on studies of the most
distant galaxies ($z \ge 6$), and a look toward the promise of a 10\%
SKA demonstrator working at short cm wavelengths for such studies.
I begin with a brief review of the current observational constraints
on cosmic reionization. For more detail, see Fan, Carilli, Keating
(2006).

\section{Observational constraints on cosmic reionization}

The last decade has seen the first observational evidence for cosmic
reionization.  The primary results come from the Gunn-Peterson effect,
ie.  Ly$\alpha$ absorption by the neutral IGM, toward the most distant
quasars ($z \sim 6$), and the large scale polarization of the CMB,
corresponding to Thomson scattering during reionization.  These
observations suggest that reionization was a complex process, with
significant variance in both space and time, starting perhaps as high
as $z \sim 14$, with the last vestiges of the the neutral IGM being
etched-away by $z \sim 6$ (see review by Fan, Carilli, Keating 2006).

Figure 1 shows the current state of observational constraints on the
cosmic neutral fraction as a function of redshift. The GP effect, and
related statistical measures of eg.  dark gaps in QSO spectra (Fan et
al. 2006), suggest a qualitative change in the nature of the IGM at $z
\sim 6$, likely signifying the end of reionization. At the other
extreme, the CMB large scale polarization suggests a significant
ionization fraction extending to $z \sim 11$.  A major task in the
coming decade is to 'connect the dots' from $z \sim 6$ out to $z \sim
14$ in this figure.

\begin{figure}
\centerline{\psfig{file=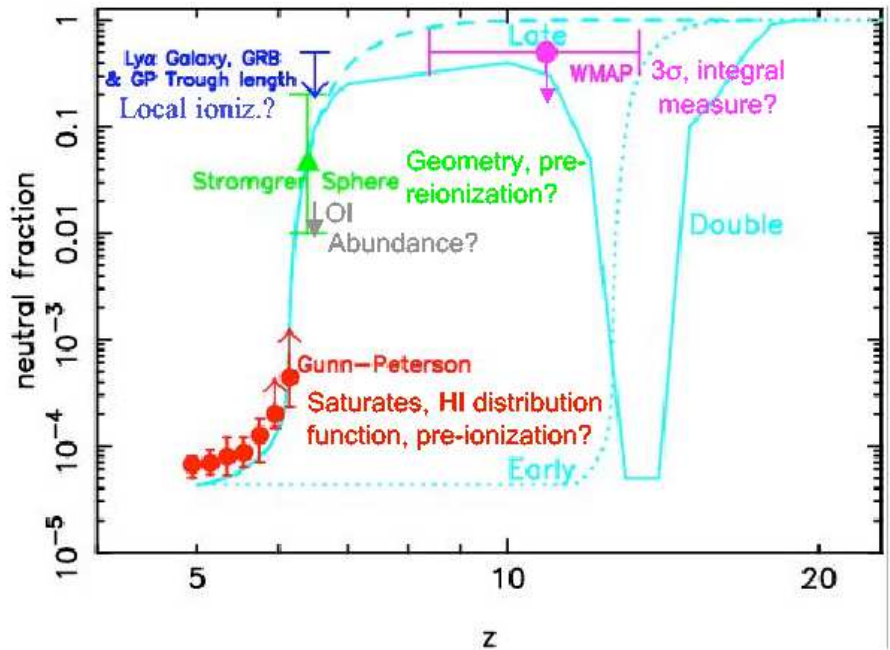,width=5in}}
\caption{The neutral fraction, by volume, of the
intergalactic medium as a function of redshift. 
Current constraints are shown, as well as representative
models for the HI evolution. Also shown are potential 
causes for uncertainty in the measurements (adapted from
Fan , Carilli, Keating 2006).
}
\label{HIsim}
\end{figure}

These first measurements of cosmic reionization are truly a water-shed
event. However, it has also become clear that these important first
probes of cosmic reionization are fundamentally limited. Figure 1 also
lists potential causes for uncertainty in the measurements (adapted
from Fan , Carilli, Keating 2006). The CMB polarization remains only a
3$\sigma$ result, and it represents an integral measure of the Thomson
scattering optical depth back to recombination, and hence can be fit
by many different reionization scenarios. For the Gunn-Peterson
effect, the IGM becomes optically thick to Ly$\alpha$ absorption for a
neutral fraction $> 10^{-3}$, and hence the diagnostic capabilities of
this probe effectively saturate at low neutral fractions.  The GP
conclusions are also depend on an assumed clumping factor for the IGM.

\section{HI 21cm signals from cosmic reionization}

\subsection{Large Scale Structure}

It is widely recognized that the most direct and incisive means of
studying cosmic reionization is through the 21cm line of neutral
Hydrogen (Furlanetto et al. 2006; Carilli 2006).  The study of HI
21cm emission from cosmic reionization entails the study of large
scale structure (LSS), meaning HI masses $> 10^{12}$ M$_\odot$. 

\begin{figure}
\centerline{\psfig{file=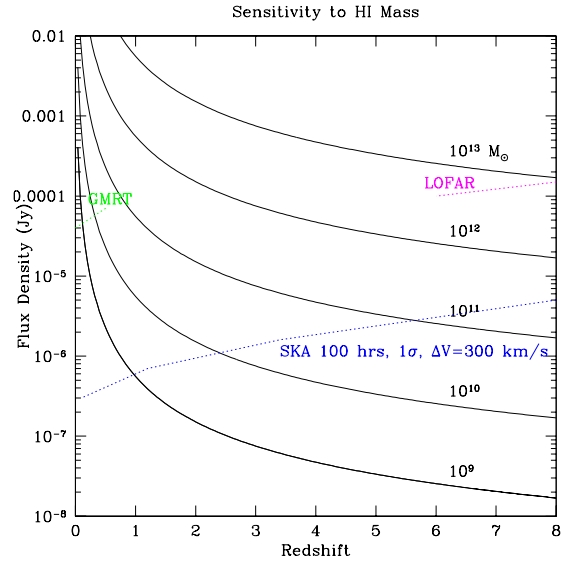,width=3in}}
\caption{The predicted HI 21cm signals versus redshift
for different HI masses, assuming a constant line width
of 300 km s$^{-1}$. Sensitivity of low frequency telescopes
are also shown, for 1000 hour integration with LOFAR and the GMRT, 
and 100 hour with SKA.  
}
\label{HIsim}
\end{figure}

Figure 2 shows the signal strength expected for a given HI mass versus
redshift. A constant line width of 300 km s$^{-1}$ was assumed for
simplicity. The key point is that, even with future large area
telescopes such as the SKA, HI measurements at these high redshifts
will be restricted to LSS, not individual galaxies.  Fortunately,
during this epoch the entire IGM can be neutral, and the LSS in
question is not simply mass clustering, but involves a combination of
structure in cosmic density, neutral fraction, and HI excitation
temperature.

The optical depth on the 21cm line of neutral hydrogen is given
by: 

\begin{equation}
\tau
 = {{3c^3 \hbar A_{10} n_{HI}}\over{16 k_B \nu_{21}^2 T_S H(z)}} \\
  \sim 0.0074 \frac{x_{HI}}{T_S} (1+\delta)(1+z)^{3/2}
  [H(z)/(\frac{dv}{dr})],
  \label{tauofz}
\end{equation}
\noindent where $A$ is the Einstein coefficient and
$\nu_{21}$ = 1420.40575 MHz (eg. Santos et al. 2005).
This equation shows immediately the rich physics involved in
studying the HI 21cm line during reionization, with $\tau$ depending
on the evolution of cosmic over-densities, $\delta$ (predominantly in
the linear regime), the neutral fraction, $x_{HI}$
(ie. reionization),  the HI excitation, or spin, temperature,
$T_S$, and the velocity structure, $\frac{dv}{dr}$, including
the Hubble flow and peculiar velocities.

In the Raleigh-Jeans limit, the observed
brightness temperature (relative to the CMB)
due to the HI 21cm line at a frequency
$\nu = \nu_{21}/(1+z)$, is given by:
\begin{equation}
T_B  \approx ~  \frac{T_S - T_{\rm CMB}}{1+z} \, \tau
\label{eq:dtb} \\
 \approx ~  7 (1+\delta) x_{HI} (1 - \frac{T_{CMB}}{T_S})
(1+z)^{1/2} ~ \rm{mK},
\end{equation}
\noindent The conversion factor from brightness temperature to
specific intensity, I$_\nu$, is given by:
$I_\nu = \frac{2 k_B}{(\lambda_{21} (1+z))^2} T_B =
22 (1+z)^{-2} T_B$ Jy deg$^{-2}$.
These equations show that for
$T_S \sim T_{CMB}$ one expects no 21cm signal.
When $T_S >> T_{CMB}$, the brightness temperature
becomes independent of spin temperature. When
$T_S << T_{CMB}$, we expect a strong negative
(ie. absorption) signal against the CMB.

The interplay between the CMB temperature, the kinetic temperature,
and the spin temperature, coupled with radiative transfer, lead to a
number of interesting physical regimes for the HI 21cm signal (Ali
2005; Barkana \& Loeb 2004; Carilli 2005; Furlanetto et al. 2006;
Fan, Carilli, Keating 2006):

\begin{itemize}

\item  At $z> 200$ equilibrium
between $T_{CMB}$, $T_K$, and T$_{S}$ is maintained by Thomson
scattering off residual free electrons and gas collisions.  In this
case $T_S = T_{CMB}$ and there is no 21cm signal.  

\item At $z \sim 30$
to 200, the gas cools adiabatically, with temperature falling as
$(1+z)^2$, ie. faster than the (1+z) for the CMB. However, the mean
density is still high enough to couple $T_S$ and $T_K$, and the HI
21cm signal would be seen in absorption against the CMB (Sethi 2005;
see section 3.5).

\item At $z \sim 20$ to 30, collisions can no longer couple $T_K$ to
$T_S$, and $T_S$ again approaches $T_{CMB}$. However, the Ly$\alpha$
photons from the first luminous objects (Pop III stars or
mini-quasars), may induce local coupling of $T_K$ and $T_S$, thereby
leading to some 21cm absorption regions (Cen 2006).  At the same time,
Xrays from these same objects could lead to local IGM warming above
$T_{CMB}$ (Chen \& Miralda-Escude 2003).  Hence one might expect a
patch-work of regions with no signal, absorption, and perhaps
emission, in the 21cm line.

\item At $z \sim
6$ to 20 all the physical processes come to play. The IGM is being
warmed by hard Xrays from the first galaxies and black holes (Loeb \&
Zaldarriaga 2004; Barkana \& Loeb 2004; Ciardi \& Madau 2003), as well
as by weak shocks associated with structure formation (Furlanetto,
Zaldarriaga, \& Hernquist 2005), such that $T_K$ is
likely larger than $T_{CMB}$ globally (Furlanetto et al. 2004b).
Likewise, these objects are reionizing the universe, leading to a
fundamental topological change in the IGM, from the linear evolution
of large scale structure, to a bubble dominated era of HII regions
(Furlanetto et al. 2005a).

\end{itemize}

\subsection{The HI 21cm signals}

{\bf Global signal:} The left panel in Figure 3 shows
predictions of the global (all sky) increase in the background
temperature due to the HI 21cm line from the neutral IGM (Gnedin \&
Shaver 2003).  The predicted HI signal peaks at roughly 20 mK above
the foreground at $z \sim 10$. At higher redshift, prior to IGM
warming, but allowing for Ly$\alpha$ emission from the first luminous
objects, the HI is seen in absorption against the CMB.  Since this is
an all sky signal, the sensitivity of the experiment is independent of
telescope collecting area, and the experiment can be done using small
area telescopes at low frequency, with very well controlled. and
calibrated, frequency response (Bowman, Rogers, Hewitt 2007).  Note
that the line signal is only $\sim 10^{-4}$ that of the mean
foreground continuum emission at $\sim 150$ MHz.

\begin{figure}
\psfig{file=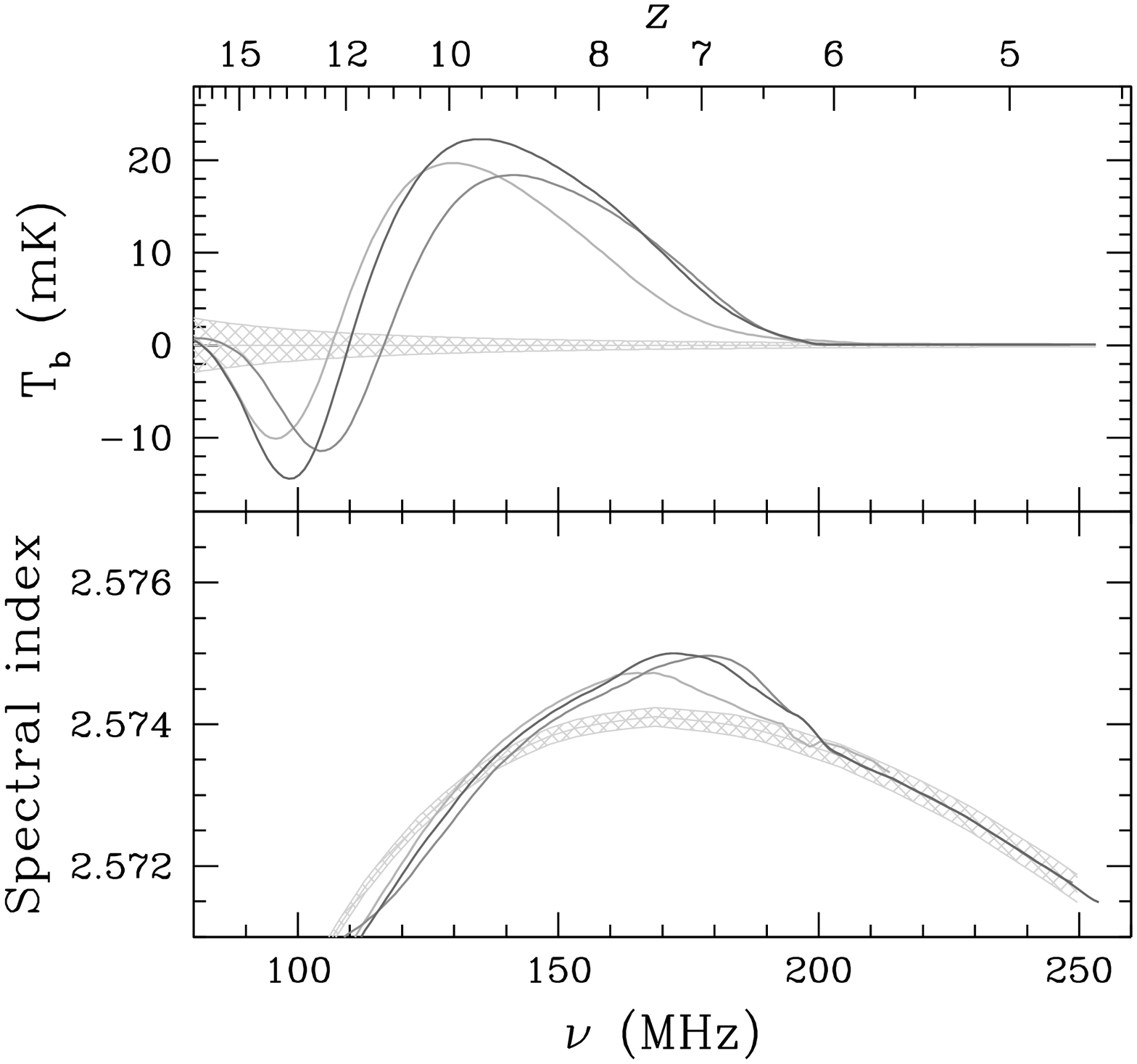,width=2.0in}
\vskip -1.8in
\hspace*{2.1in}
\psfig{file=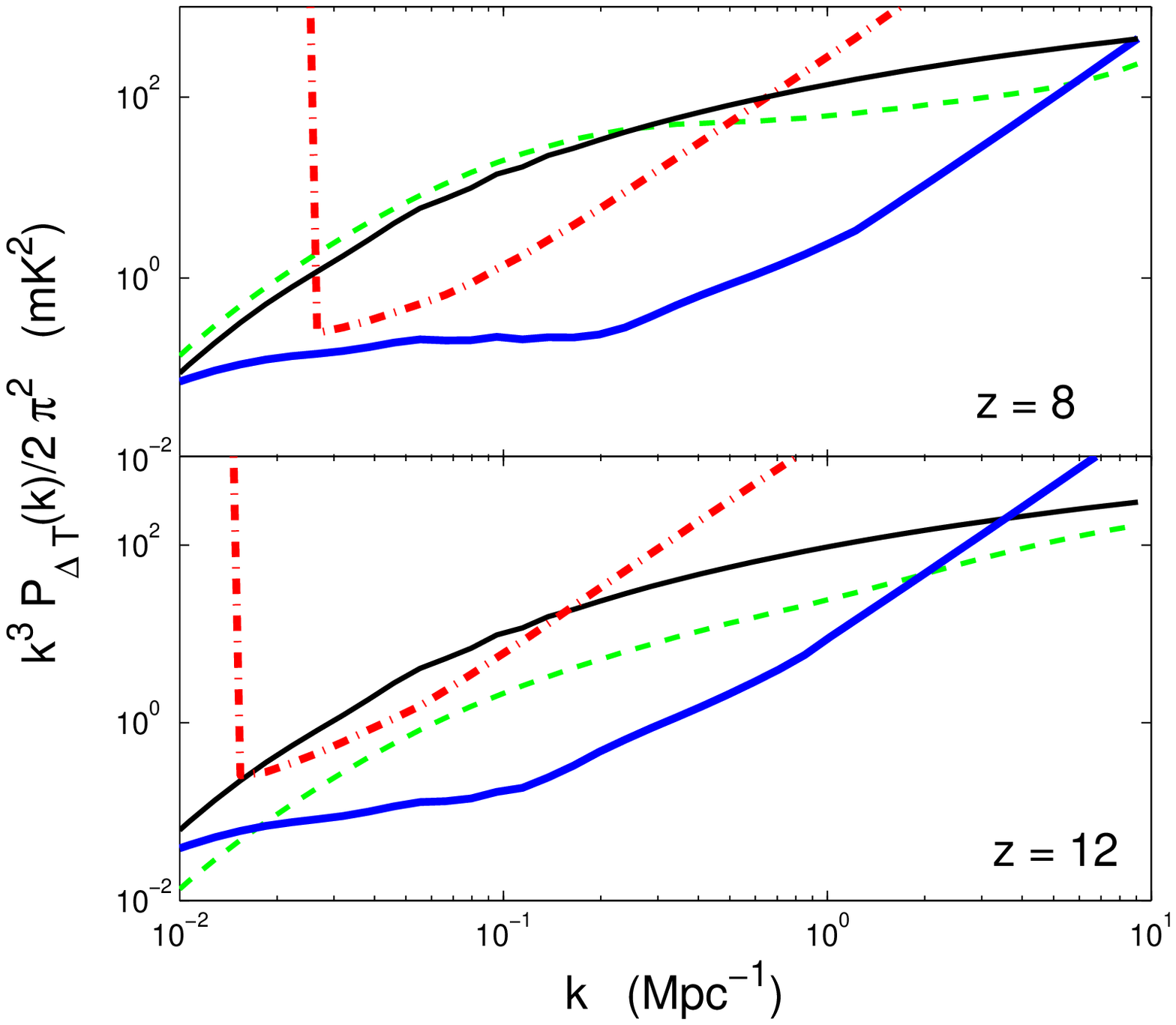,width=2.0in}
\vskip -1.82in
\hspace*{4.2in}
\psfig{file=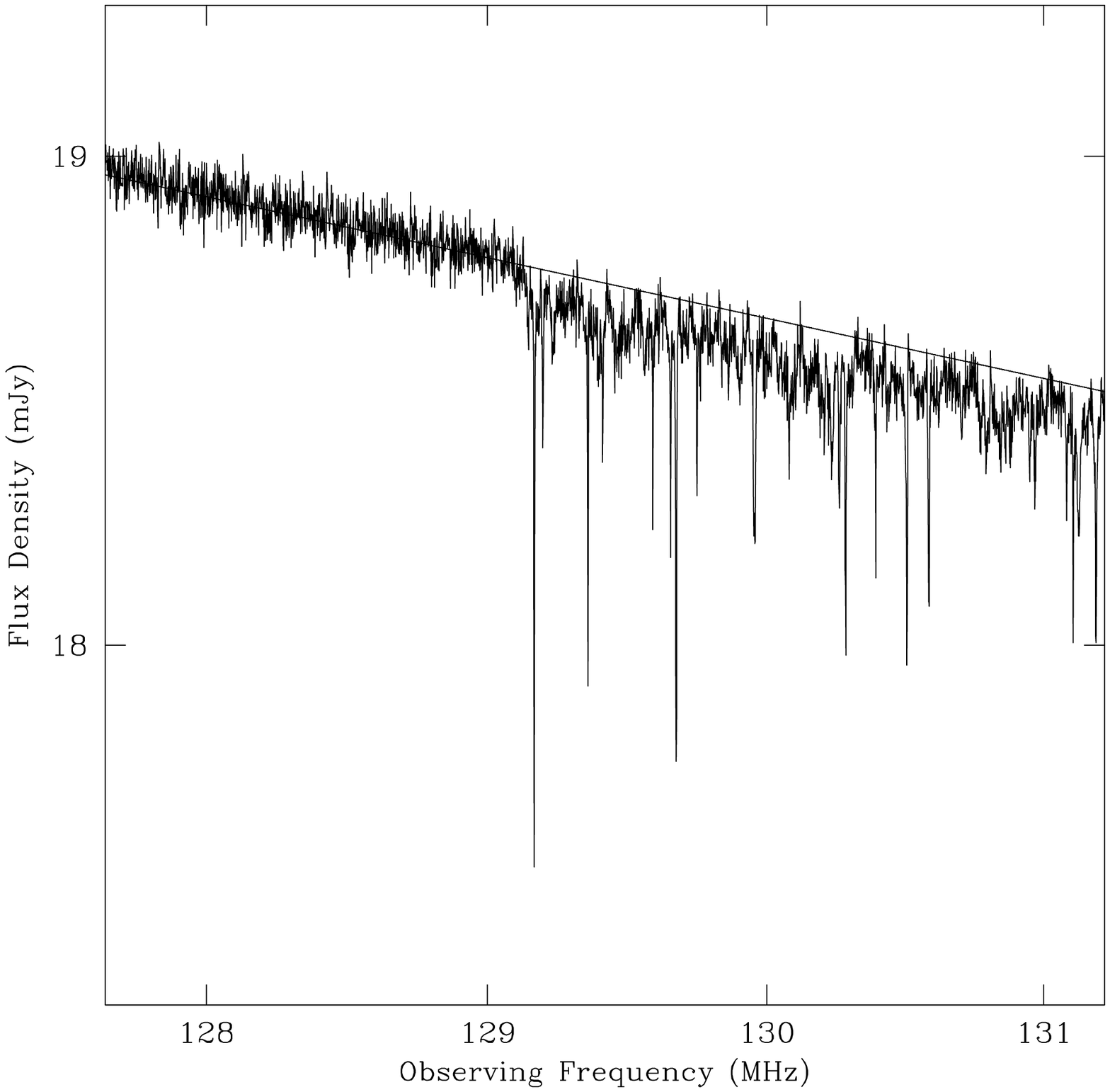,width=1.8in}
\hskip 0.2in
\caption{{\bf Left}: Global (all sky) HI signal from reionization
(Gnedin \& Shaver 2003). 
The shaded region shows the expected thermal
noise in a carefully controlled experiment.
{\bf Center:} Predicted HI 21cm brightness temperature power spectrum
 (in log bins) at redshifts 8 and 12 
(Mcquinn et al. 2006).   The thin
black line shows the signal when density fluctuations dominate.  The
dashed green line shows the predicted signal for $\bar{x}_i = 0.2$
at $z=12$, and $\bar{x}_i = 0.6$ at $z=8$,
in the (Furlanetto et al. 2004) semi-analytic model. 
The thick blue line
shows the SKA sensitivity in 1000hrs. The thick red dot-dash show the
sensitivity of the pathfinder experiment LOFAR. The cutoff at low k is
set by the primary beam.
{\bf Right}:  The simulated SKA spectrum of a radio continuum source
at $z = 10$ (Carilli et al. 2002). 
The straight line is the intrinsic
power law (synchrotron) spectrum of the source. The noise curve
shows the effect of the 21cm line in the neutral IGM, including
noise expected for the SKA in a 100 hour integration.}
\end{figure}

{\bf Power spectra:} The middle panel in Figure 3 shows
the predicted power spectrum of spatial fluctuations in the sky
brightness temperature due to the HI 21cm line (Mcquinn et al. 2006).
For power spectral analyses the sensitivity is greatly enhanced
relative to direct imaging due to the fact that the universe is
isotropic, and hence one can average the measurements in annuli in the
Fourier (uv) domain, ie. the statistics of fluctuations along an
annulus in the uv-plane are equivalent.  Moreover, unlike the CMB, HI
line studies provide spatial and redshift information, and hence the
power spectral analysis can be performed in three dimensions. The rms
fluctuations at $z = 10$ peak at about 10 mK rms on scales $\ell \sim
5000$.

A recent analysis by Lidz et al. (2008) shows that the MWA should be
able to determine the HI 21cm power-spectrum over roughly a decade in
wavenumber, $k \sim 0.1$ to 1$h$ Mpc$^{-1}$.  These data should be
able to constrain both the amplitude and slope of the power
spectrum. A key aspect of these measurements is determination of the
redshift evolution of these quantities, with the amplitude of the rms
power in the HI 21cm fluctuations peaking at redshift when the IGM
is roughly 50\% neutral.

{\bf Absorption toward discrete radio sources:} An interesting
alternative to emission studies is the possibility of studying smaller
scale structure in the neutral IGM by looking for HI 21cm absorption
toward the first radio-loud objects (AGN, star forming galaxies, GRBs)
(Carilli et al. 2002). The rightpanel of Figure 3 shows the predicted
HI 21cm absorption signal toward a high redshift radio source due to
the `cosmic web' prior to reionization, based on numerical
simulations.  For a source at $z = 10$, these simulations predict an
average optical depth due to 21cm absorption of about 1$\%$,
corresponding to the `radio Gunn-Peterson effect', and about five
narrow (few km/s) absorption lines per MHz with optical depths of a
few to 10$\%$. These latter lines are equivalent to the Ly $\alpha$
forest seen after reionization (a recent treatment of this problem can
be found in Furlanetto 2006). Furlanetto \& Loeb (2002) predict a
similar HI 21cm absorption line density due to gas in minihalos as
that expected for the 21cm forest.

{\bf Tomography:} Figure 4 shows the expected evolution of the HI 21cm
signal during reionization based on numerical simulations (Zaldariagga
et al. 2004; see also Zahn et al. 2007; Iliev et al. 2006; Shapiro et
al. 2006, etc...).  In this simulation, the HII regions caused by
galaxy formation are seen in the redshift range $z \sim 8$ to 10,
reaching scales up to 2$'$ (frequency widths $\sim 0.3$ MHz $\sim 0.5$
Mpc physical size). These regions have (negative) brightness
temperatures up to 20 mK relative to the mean HI signal.  This
corresponds to 5$\mu$Jy beam$^{-1}$ in a 2$'$ beam at 140 MHz. Only
the full SKA will able to image such structures.

\begin{figure}
\centerline{\psfig{file=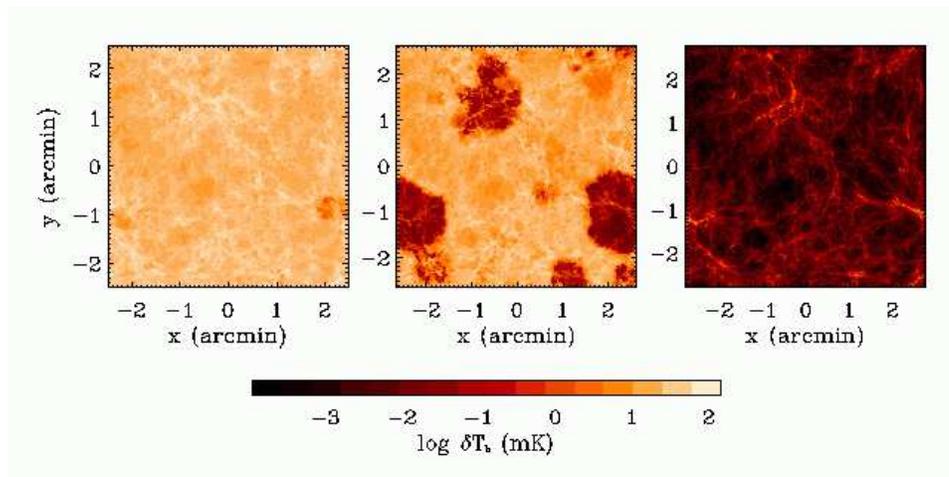,width=5in}}
\caption{The simulated HI 21cm brightness temperature distribution
during reionization at $z = 12$, 9, 7 (left to right)
(Zaldariagga et al. 2004)}
\label{HIsim}
\end{figure}

{\bf Largest Cosmic Stromgren Spheres (CSS):} While direct detection
of the typical structure of HI and HII regions may be out of reach of
the near-term EoR 21cm telescopes, there is a chance that even this
first generation of telescopes will be able to detect the rare, very
large scale HII regions associated with luminous quasars near the end
of reionization.  The expected signal is $\sim 20$mK $\times x_{HI}$
on a scale $\sim 10'$ to 15$'$, with a line width of $\sim 1$ to 2 MHz
(Wyithe, Loeb, Barnes 2005).  This corresponds to 0.5 $\times x_{HI}$
mJy beam$^{-1}$, for a 15$'$ beam at $z \sim 6$ to 7, where $x_{HI}$
is the IGM neutral fraction. Figure 5 shows a simulation of the
expected images from near-term reionization telescopes, such as the
MWA, for these very large structures. The key aspect of these systems
is that the structures are three dimensional, making them much easier
to detect than continuum structures at a similar brightness.

Wyithe et al. (2005) predict that, for late reionization, there should
be several tens of (mostly fossil) large ($> 4$Mpc physical) CSS per
10$^o$ field of view in $z \sim 6$ to 8.  An interesting addition to
the QSO CSS studies is the possibility that massive galaxy formation
at very high redshift is highly clustered, occuring in rare peaks in
the cosmic density field.  The total number of ionizing photons from
star formation integrated over the lifetime of the system can match
that radiated by a luminous QSO (Li et al. 2007; Wyithe et al. 2007),
and hence generate CSS comparable to those predicted for the QSOs. The
star forming systems will both increase the size of the spheres around
bright QSOs (since QSO and massive galaxy formation are likely to be
coupled), as well as increase the number of large spheres, in cases
where the QSO has not turned-on yet.

\begin{figure} 
\centerline{\psfig{file=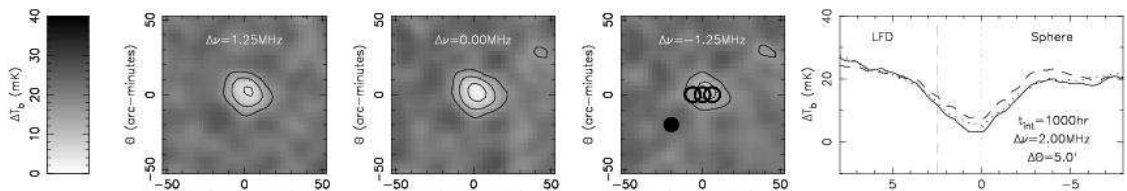,width=6in}} 
\caption{The
simulated HI 21cm brightness temperature distribution for the largest
expected structures during reionization, those associated with
luminous QSOs, or highly clustered massive galaxy formation (Wyithe,
Loeb, Barnes 2005). Left shows channel images with 1.25 MHz width,
and right shows the spectrum at the position of the source.  }
\end{figure}

To add a personal bias to this review, my prediction is that the first
robust detection of the neutral IGM using the 21cm line will not be
through power-spectral studies, but come through a repeatable, very
large, three dimensional 'hole' seen in images made in the very wide,
deep field ($\Delta \theta \ge 10^o$, $\Delta z \ge 2$), with
up-coming reionization telescopes.

\subsection{Telescopes}

Table 1 summarizes the current experiments under construction to study
the HI 21cm signal from cosmic reionization.  These experiments vary
from single dipole antennas to study the all-sky signal, to 10,000
dipole arrays to perform the power spectral analysis, and potentially
to image the largest structures during reionization (eg. the quasar
Stromgren spheres).

Most of the experiments have a few to 10\% of the collecting area of
the SKA, and there are many common features:

\begin{itemize}

\item All rely on
some form of a wide-band dipole or spiral antenna, eg.  log-periodic
yagis, sleeve dipoles, or bow-ties, with steering of the array
response through electronic phasing of the elements.  

\item The
front-end electronics are relatively simple (uncooled
amplifier/balun), since the system performance is dominated by the sky
brightness temperature. 

\item Most rely on a grouping of dipoles
into 'tiles' or 'stations', to decrease the field-of-view, and to
decrease the data rate into the correlator to a managable level.  

\item The large number of array elements, and the need for wide-field,
high dynamic range imaging over an octave, or more, of bandwidth,
demand major computing resources, both for basic cross-correlation,
and subsequent calibration, imaging, and analysis. For example,
assuming an array of 1000 tiles, a 100 MHz bandwidth, and 8 bit
sampling, the total data rate coming into the correlator is 1.6 Tbit
s$^{-1}$. The LOFAR array is working with IBM to apply the 27.4 Tflop
Blue Gene supercomputing technology to interferometric imaging (Falcke
2006).

\end{itemize}

First results from these path-finder telescopes are expected within
the next few years.  Experience from these observations in dealing
with the interference, ionosphere, and wide-field imaging/dynamic
range problems will provide critical information for future
experiments, such as the Square Kilometer Array, or a low frequency
radio telescope on the moon.

\begin{figure}[!t]
\psfig{file=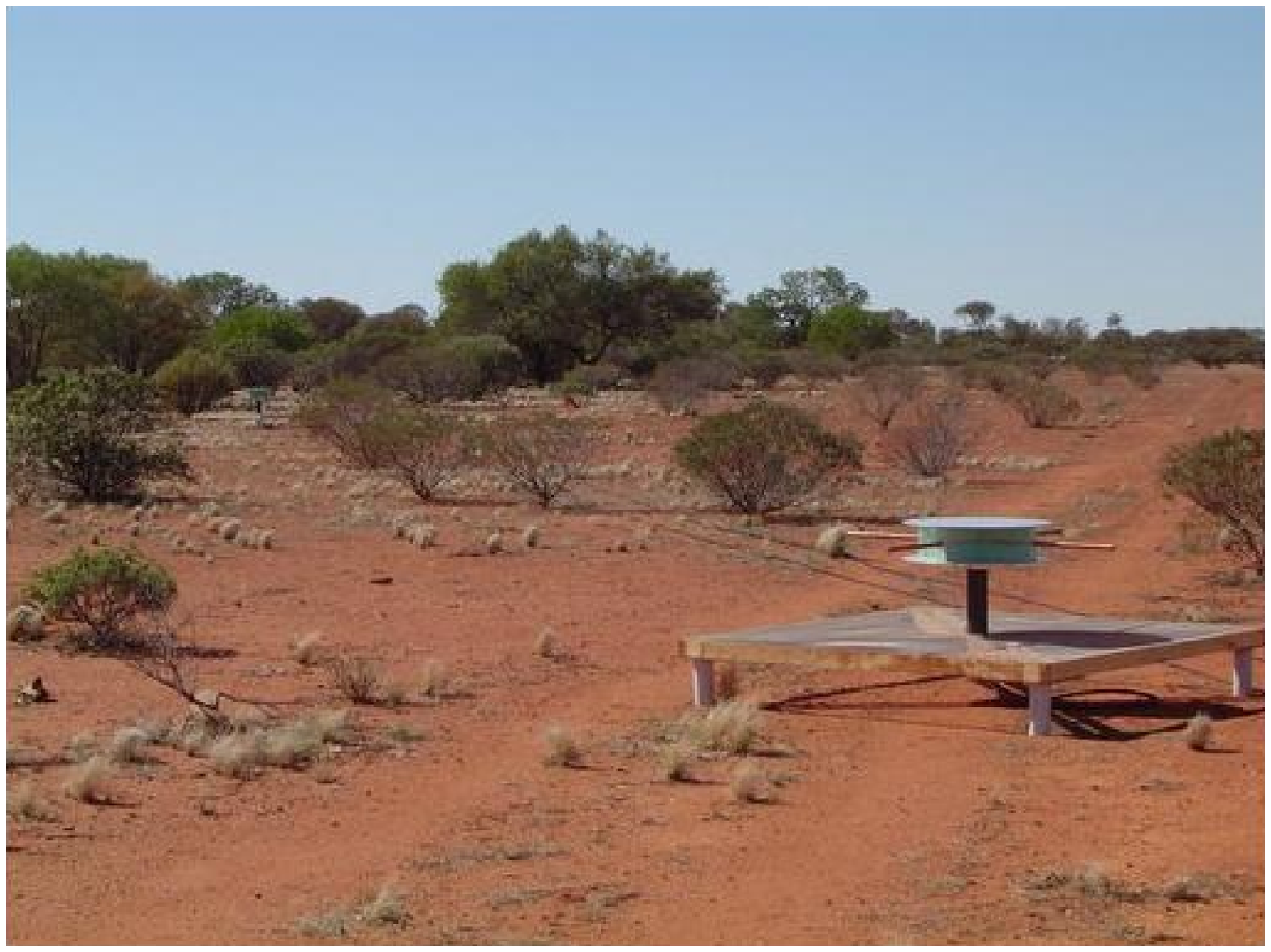,width=3.0in}
\vskip -2.5in
\hspace*{3.4in}
\psfig{file=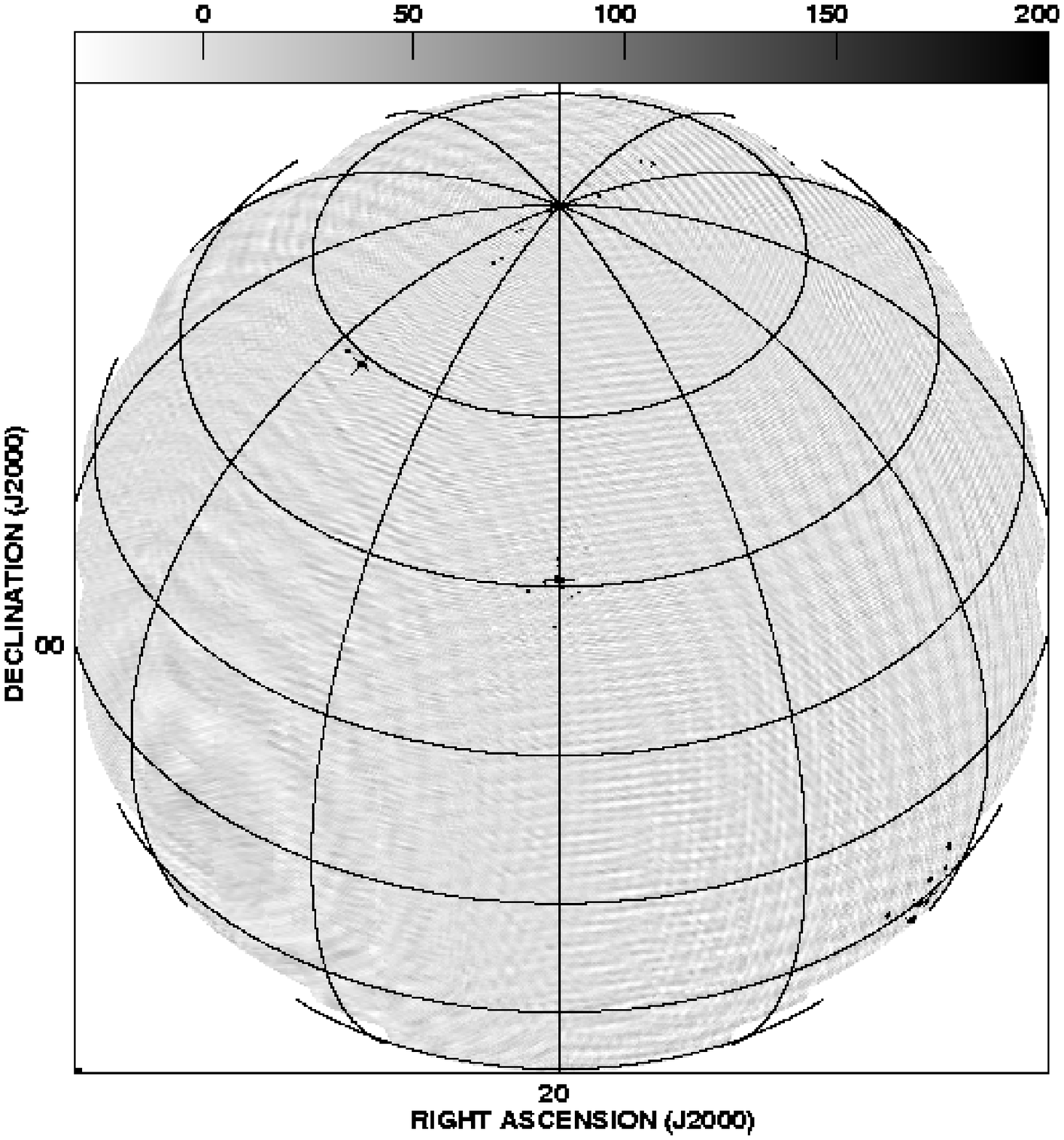,width=2.5 in}
\caption{Left: A few elements of the PAPER test array in Western
Australia (Mastrantonio \& Backer 2007 in prep). 
Right:  A full sky image from PAPER-GB
showing the brightest sources (Cas A, Cygnus A, Sun).}
\end{figure}

As an example, Figure 6 shows dipole elements and an all sky image
from the Precision Array to Prope the Epoch of Reionization (PAPER).
This array is being built in Western Australia, with a test
installation in Green Bank, with the purpose of performing near-term
studies of the HI 21cm signal from reionization. 

\begin{table}
  \begin{center}
  \begin{tabular}{lcccccc}
\hline
Experiment & Site & Type & $\nu$ range & Area & Date & Goal \\
~ & ~ & ~ & MHz & m$^2$ & ~ & ~ \\
\hline
\hline
Mark I$^a$ & Australia & spiral & 100-200 & few & 2007 & All Sky \\
EDGES$^b$ & Australia & four-point & 100-200 & few & 2007 & All Sky \\
GMRT$^c$ & India & parabola array & 150-165 & 4e4 & 2007 & CSS$^d$ \\
PAPER$^e$ & Australia & dipole array & 110-190 & 1e3 & 2008 & PS/CSS/Abs \\
21CMA$^f$ & China & dipole array & 70-200 & 1e4 & 2007 & PS \\
MWAd$^g$  & Australia & aperture array & 80-300 & 1e4 & 2008 & PS/CSS/Abs \\
LOFAR$^h$ & Netherlands & aperture array & 115-240 & 1e5 & 2008 & PS/CSS/Abs \\
SKA$^i$ & ? & aperture array & 100-200 & 1e6 & 2015 & Imaging \\
\hline
  \end{tabular}
  \caption{HI 21cm Cosmic Reionization Experiments}
  \label{tab:tab}
  \end{center}
$^a$http://www.atnf.csiro.au/news/newsletter/jun05/Cosmological\_re-ionization.htm \\
$^b$http://www.haystack.mit.edu/ast/arrays/Edges/index.html \\
$^c$http://gmrt.ncra.tifr.res.in/ \\
$^d$CSS = Cosmic Stromgren Spheres, PS = power spectrum, Abs = absorption \\
$^e$http://astro.berkeley.edu/$\sim$dbacker/EoR/ \\
$^f$http://cosmo.bao.ac.cn/project.html \\
$^g$http://www.haystack.mit.edu/ast/arrays/mwa/ \\
$^h$http://www.lofar.org/ \\
$^i$http://www.skatelescope.org/ \\
\end{table}

\subsection{Observational challenges to HI 21cm studies of cosmic
reionization}

{\bf Foregrounds:} The HI 21cm signal from reionization must be
detected on top of a much larger synchrotron signal from foreground
emission. This foreground includes discrete radio galaxies,
and large scale emission from our own Galaxy. The expected HI
21cm signal is about $10^{-4}$ of the foreground emission at 140 MHz.

di Matteo et al. (2002) show that, even if point sources can be
removed to the level of 1 $\mu$Jy, the rms fluctutions on spatial
scales $\le 10'$ ($\ell \ge 1000$) due to residual radio point sources
will be $\ge 10$ mK just due to Poisson noise, increasing by a factor
100 if the sources are strongly clustered (see also di Matteo, Ciardi,
\& Miniati 2004; Oh \& Mack 2003). 

A key point is that the foreground emission should be smooth in
frequency, predominantly the sum of power-law, or perhaps gently
curving, non-thermal spectra.  A number of complimentary approaches
have been presented for foreground removal (Morales, Bowman, \& Hewitt
2005).  Gnedin \& Shaver (2003) and Wang et al (2005) consider fitting
smooth spectral models (power laws or low order polynomials in log
space) to the observed visibilities or images. Morales \& Hewitt
(2003) and Morales (2004) present a 3D Fourier analysis of the
measured visibilities, where the third dimension is frequency.  The
different symmetries in this 3D space for the signal arising from the
noise-like HI emission, versus the smooth (in frequency) foreground
emission, can be a powerful means of differentiating between
foreground emission and the EoR line signal.  Santos et al. (2005),
Bharadwaj \& Ali (2005), Zaldariagga et al. (2004a) perform a similar
analysis, only in the complementary Fourier space, meaning cross
correlation of spectral channels. They show that the 21cm signal will
effectively decorrelate for channel separations $> 1$ MHz, while the
foregrounds do not.  The overall conclusion of these methods is that
spectral decomposition should be adequate to separate synchrotron
foregrounds from the HI 21cm signal from reionization at the mK level.

Glesser, Nusser, \& Bensor (2008) recently considered the relative
contribution of various contributions to the mean sky brightness
temperature at low frequency. They show that roughly 90\% comes from
the Galactic synchrotron emission, close to 10\% from extragalactic
radio sources, and up to $\sim 2\%$ from Galactic free-free emission.
They also show that these foregrounds can be removed effectively
through image reconstruction techniques that take into consideration
the very different frequency structure of the cosmological line signal
and and the foreground continuum emission.

{\bf Ionosphere:} A second potential challenge to low frequency
imaging over wide fields is phase fluctuations caused by the
ionosphere.  These fluctuations are due to index of refraction
fluctuations in the ionized plasma, and behave as $\Delta \phi \propto
\nu^{-2}$.  

Figure 7 shows an example of ionospheric phase errors on source
positions using VLA data at 74 MHz (Cotton et al. 2004; Lane et
al. 2004).  Positions of five sources are shown for a series of snap
shot images over 10 hours.  A number of interesting phenomena can be
seen. First, the sources are slowly moving in position over time, by
$\pm 50"$ over timescales of hours. These position shifts reflect the
changing electronic path-length due to the fluctuating ionosphere (ie.
tilts in the incoming wavefront due to propagation delay).  Second,
the individual sources move roughly independently. This is a
demonstration of the 'isoplanatic patch' problem, ie.  the excess
electrical path-length is different in different directions. At 74
MHz, the typical coherent patch size is about 3$^o$ to
4$^o$. Celestial calibrators further than this distance from a target
source no longer give a valid solution for the combined instrumental
and propagation delay term required to image the target source.  And
third, at the end of the observation there occurs an ionospheric
storm, or traveling ionospheric disturbance, which effectively
precludes coherent imaging during the event.

\begin{figure}
\centerline{\psfig{figure=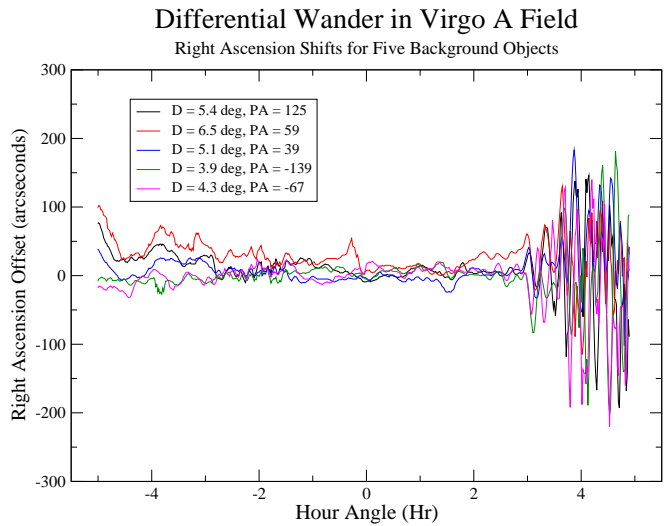,width=4in,angle=-90}}
\caption{The positions of five sources in the Virgo A field at 74 MHz
observed with the VLA over 10 hours (Cotton et al. 2004; Lane et
al. 2004). The source positions vary with time due to fluctuations in
the electronic path-length through the ionosphere.  } \label{}
\end{figure}

New wide field self-calibration techniques, involving multiple phase
solutions over the field, or a `rubber screen' phase model (Cotton et
al. 2004), are being developed that should allow
for self-calibration over wide fields.

{\bf Interference:}
Perhaps the most difficult problem facing low frequency radio
astronomy is terrestrial (man-made) interference (RFI). The relevant
frequency range corresponds to 7 to 200 MHz ($z= 200$ to 6). These are
not protected frequency bands, and commercial allocations include
everything from broadcast radio and television, to fixed and mobile
communications.

Many groups are pursuing methods for RFI mitigation and excision (see
Ellingson 2004). These include: (i) using a reference horn, or one
beam of a phased array, for constant monitoring of known, strong, RFI
signals, (ii) conversely, arranging interferometric phases to produce
a null at the position of the RFI source, and (iii) real-time RFI
excision using advanced filtering techniques in time and frequency, of
digitized signals both pre- and post-correlation. The latter requires
very high dynamic range (many bit sampling), and very high frequency
and time resolution.

In the end, the most effective means of reducing interference is to go
to the remotest sites.  Figure 8 shows the interference measured at
three different locations with population densities changing by
roughly four orders of magnitude (Chippendale \& Beresford 2006 ATNF
RFI monitoring program).  The MWA, PAPER, and PAST have selected sites
in remote regions of Western Australia, and China, because of known
low RFI environments. Of course, the ultimate location would be the
far-side of the moon (section 3.5).

\begin{figure}
\centerline{\psfig{figure=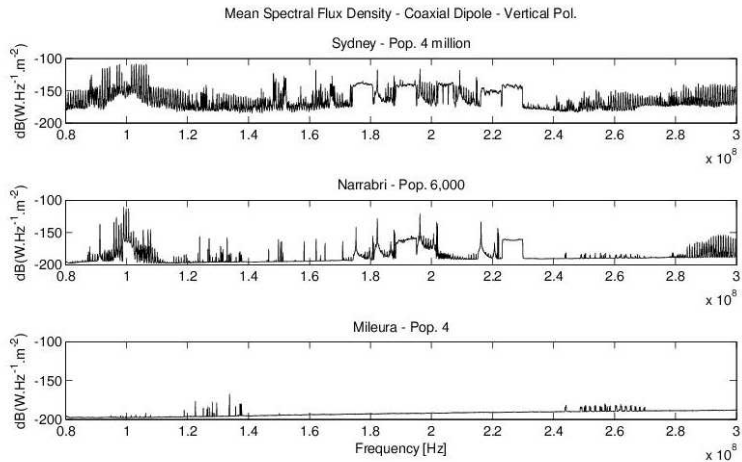,width=4in,angle=-90}}
\caption{The interference environment in
Australia. The FM band is between 87.5 and 108 MHz.
The TV stations are between 180 and 200MHz,
including broad band digital transmissions.
Shown courtesy of A. Chippendale and R. Beresford (taken as part of the
ATNF SKA Site Monitoring Program.)}
\end{figure}

The technical challenges to HI 21cm observations of reionization are
many. Use of spectral decomposition to remove the foregrounds requires
careful control of the synthesize beam as a function of frequency,
with the optimal solution being a telescope design where the
synthesized beam is invariant as a function of frequency (Bowman et
al. 2007).  High dynamic range front ends are required to avoid
saturation in cases of strong interference, while fine spectral
sampling is required to avoid Gibbs ringing in the spectral response.
The polarization response must be stable and well calibrated to remove
polarized foregrounds (Haverkorn et al. 2004). Calibation in the
presence of a structured ionospheric phase screen requires new wide
field calibration techniques.  The very high data rate expected for
many element ($\ge 10^3$) arrays requires new methods for data
transmission, cross correlation, and storage.

\subsection{The lunatic fringe: A PAPER Moon}

At the lowest frequencies, $\le 50$MHz or so, where we hope to study
the Pre-galactic medium (PGM), phase fluctuations and the opacity of
the ionosphere become problematic. There is long standing interest in
building a low frequency radio telescope on the far side of the moon
(Gorgolewski 1965; Burke 1985; Kuiper et al. 1990; Burns \& Asbell
1991; Woan et al. 1997). The reasons are clear (section 2): no
ionosphere, and sheilding from terrestrial radio frequency
interference (RFI). Two factors have acted to rekindle interest in low
frequency radio astronomy from moon. First, scientifically, efforts to
study the PGM using the redshifted HI 21cm line have spurred numerous
ground-based low frequency projects (section 3).  And second is the
new NASA initiative to return Man to the moon, and beyond.

While not a scientific rationale, it should be pointed out that a low
frequency telescope may be the easiest astronomical facility to deploy
and maintain on the moon. The antennas and electronics are high
tolerance, with wavelengths $> 1.5$m and system noise characteristics
dominated by the Galactic foreground radiation. Deployment could be
automated, using either javelin deployment (EADS/ASTRON), rollout of
thin polyimide films with metalic deposits (ROLSS; Lazio et al. 2006),
inflatable dipoles, or deployment by rovers. Likewise, being a phased
array, low frequency telescopes are electronically steered, and hence
have no moving parts. Lastly, there is no potential difficulty with
lunar dust affecting the optics.

Figure 9 shows radio power received by a low frequency radio receiver
on a lunar orbiter from the 1970's (Alexander et al. 1975).  The
strong power received from Earth's auroral emission is completely gone
during immersion.

\begin{figure}
\centerline{\psfig{figure=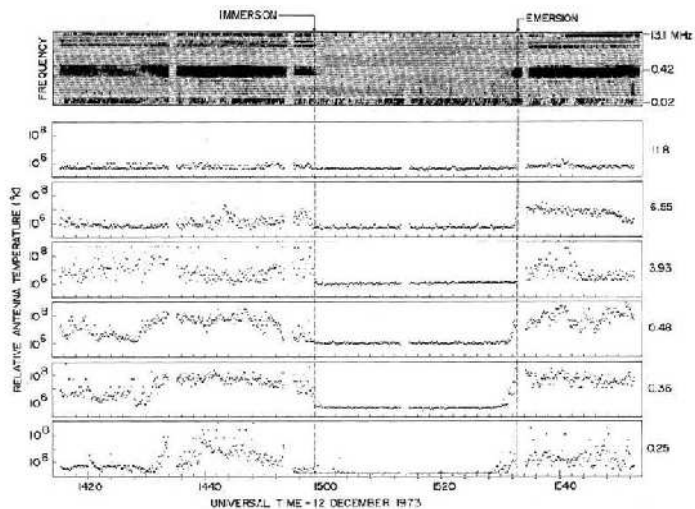,width=4in}}
\caption{Radio power received by the lunar orbiting radio
astronomy explorer satellite in 1973 (RAE2; Alexander et al. 1975).
The power is dominated by the Earth's auroral emission except 
during immersion, when the Earth is totally blocked by the moon.
} \label{}
\end{figure}

At the very low frequencies corresponding to the dark ages, going to
the Moon becomes imperative, due to the rapidly increasing ionospheric
opacity and phase effects.  A number of studies have considered this
very low frequency, pre-reionization HI 21cm signal (Loeb \&
Zaldarriaga 2004; Cen 2006; Barkana \& Loeb 2005b; Shethi 2005).  The
HI 21cm measurements can explore this physical regime at $z \sim 50$
to 300, or redshifts $z > 30$.

In this redshift regime the HI generally follows linear density
fluctuations, and hence the experiments are as clean as CMB studies,
and $T_K < T_{CMB}$, so a relatively strong absorption signal might be
expected.  Also, Silk damping, or photon diffusion, erases structures
on scales $\ell > 2000$ in the CMB at recombination, corresponding to
comoving scales = 22 Mpc, leaving the 21cm studies as the only
current method capable of probing to very large $\ell$ in the
linear regime. The predicted rms brightness temperature
fluctuations are 1 to 10 mK on scales $\ell= 10^3$ to 10$^6$ (0.2$^o$
to 1$''$).  These observations could provide the best test of
non-Gaussianity of density fluctuations, and constrain the running
power law index of mass fluctutions to large $\ell$, providing
important tests of inflationary structure formation. Sethi (2005) also
suggests that a large global signal, up to -0.05 K, might be expected
for this redshift range.

The difficulty in this case is one of sensitivity. The sky temperature
is $> 10^4$K, and using the equations in section 3.1, it is easy to
show that, even if structures as bright at 10mK on scales of a few
arcminutes exist, it would require $\sim 10$ square kilometers of
collecting area to detect them. For the more typical small scale
structure being considered (ie. $\sim 10"$), the required collecting
area increases to $3.6\times 10^{10}$ m$^2$. The dynamic range
requirements also become extreme, $> 10^8$.  It should also be kept in
mind that the sky becomes highly scattered due to propagation through
interstellar and interplanetary plasma, with source sizes obeying:
$\rm \theta_{min} \sim 1 ({{\nu}\over{1 MHz}})^{-2}$deg. Hence, all
objects in the sky are smeared to $> 4"$ for frequencies $< 30$ MHz.

\subsection{The HI n=2 fine structure line: An another probe
of the neurtal IGM?}

Recently, Sethi et al. (2007) have proposed alternative study of the
HI $2s_{1/2}-2p_{3/2}$ transition ({\it in absorption}), with 9911 MHz
rest frequency.\footnote{Transitions among hyperfine components are at
9852, 9876, and 10030 MHz in a ratio of 1:5:2; Ershov 1987.}  This
transition has been notoriously difficult to detect in the nearby
universe for two principal reasons related to the (large) Ly$\alpha$
Einstein A coefficient, $\sim 10^8$ s$^{-1}$ (Dennison et
al. 2005).  First, only $\sim 10^{-18}$ of atoms will be
in the excited $2p$ state, since the lifetime is only a few
nanoseconds; emission of a radio photon and decay to the $2s$ state is
improbable. Second, the uncertainty principle dictates a $\sim
100$\,MHz line width, due to the characteristic time scale for decay.
Detecting such weak, broad lines is difficult.  Additional factors
relevant to the nearby universe include local dust absorption of
Ly$\alpha$ photons that would otherwise pump the $2p$ state and
collisional coupling of $2s$ and $2p$ in dense regions.

Sethi et al. argue that the situation for the pristine neutral IGM at
high redshift may be different.  Gunn-Peterson absorption toward
quasars at $z > 6$ suggests that they are surrounded by ``Cosmological
Str\"omgren Spheres'' (CSS), on physical scales of $\sim 5$\,Mpc or
$\sim 15'$ (Fan et al. 2006).  The IGM outside the
CSS could still be substantially neutral and relatively unperturbed.
Non-ionizing UV photons between Ly$\alpha$ and the Lyman limit would
propagate into the neutral medium, be absorbed, and be re-emitted as
line radiation (e.g., resonant scattering of Ly$\alpha$; degradation
of Ly$\beta$ into H$\alpha$; two-photon decay $2s-1s$).  Sethi et
al. show that through line scattering processes a tiny population in
the metastable 2s state (A $\sim 8$ s$^{-1}$) can be maintained $\sim
10^8$ times longer than without the photon field excited by the
quasar.  Unlike the 21-cm transition, the fine structure transition
would be seen in {\it absorption} against the microwave background.

Sethi et al. perform a detailed calculation of the expected brightness
of the $2s_{1/2}-2p_{3/2}$ transition.  For an assumed UV radiation
field and physical conditions in the IGM just outside a CSS around a
$z\sim 6$ quasar, they predict a brightness temperature of $T_B \sim
-20 f_{HI} ~ \mu$K.  The prospect of 9911 MHz emission is promising
for late reionization because it would be redshifted into the 1.4 GHz
band, where existing dish antennas, and optimized feed and receiver
systems, are available.  In contrast, study of the redshifted HI 21cm
line requires new facilities, VHF operation, and low-gain dipole
feeds. Most importantly, the foregrounds are dramatically reduced
at higher frequency. 

There are some significant uncertainties in these calculations.
The Ly$\alpha$ rate of 10$^{58}$\,s$^{-1}$ assumed by
Sethi et al. is $\sim 10\times$ higher than has been commonly
estimated for high redshift quasars (e.g., Wyithe, Loeb, \& Carilli
2005; Fan et al. 2006).  However, re-estimation of
statistical weights and consideration of other Lyman series
transitions (Ly$\gamma$ and above) balances revision in luminosity.
Additional uncertainty surrounds the possibility of UV-absorbing dust
in the early IGM due to early stellar processing and expolusion.   

A straight-forward calculation shows that existing instruments such as
the GBT, Parkes, and the most compact configuration of the ATNF, could
potentially detect, or set interesting limits on, the neutral fraction
of the IGM at $z\sim 6$ in reasonable integration times $\sim 100$
hours.  The primary challenge in detecting $2s_{1/2}-2p_{3/2}$
absorption will be achieving the very high spectral dynamic range
required, of order 10$^6$.

\section{Radio studies of the first galaxies}

\subsection{Introduction}

The last few years has seen remarkable advances in the discovery of
luminous objects into cosmic reionization, ie. $z \ge 6$. This
includes over 100 star forming galaxies discovered through the
Ly-break and Ly$\alpha$, and related, techniques (Ellis 2006), and
some tens of QSOs (Fan et al. 2006).  An important point to keep in
mind when observing the first galaxies is that the GP effect precludes
detection of these sources at observing wavelengths $\le
900$nm. Hence, study of the first luminous objects is the regime of
near-IR through radio astronopy, and hard X-rays.

Radio astronomy has also detected objects at $z \sim 6$ at cm and mm
wavelengths in line and continuum emission.  In this section I want to
emphasize complementarity.  Proper study of the complex process of
galaxy formation requires observations across the electromagnetic
spectrum.  Near-IR observations reveal the ionized gas, stars, and
AGN.  Far IR through (sub)mm observations reveal the dust, higher
order molecular line emission, and low excitation fine structure line
emission. The higher cm frequencies reveal the lower order molecular
transitions, which provide the best estimates of total gas mass. And
at longer cm wavelengths one observes synchrotron emission, which is a
dust-unbiased measure of the star formation rate and distribution, and
of obscured AGN.  The mm and cm studies are crucial to study the cool
thermal material, the fuel for galaxy formation, and to probe galaxies
in their earliest, dust-enshrouded stages of formation.

In this section I also emphasize the fact that a full SKA is not
absolutely required for this program, and that a 10\% Phase I SKA
operating up to $\sim 40$ GHz would be a major step toward meeting
this KSP (Carilli 2006).

\subsection{The case of 1148+5251 at $z =6.42$}

To illustrate the power of our radio studies, we describe in some
detail the particularly enlightening example of the most distant SDSS
quasar, J1148+5251 at $z=6.419$, a luminous broad line AGN with a
black hole mass $\sim 10^9$ M$_\odot$ (Fan et al. 2004).  The host
galaxy has been detected in thermal dust, and non-thermal radio
continuum, as well as molecular line emission, including multiple CO
transitions, and interesting limits to dense molecular gas tracers,
such as HCN (Walter et al. 2003; Bertoldi et al. 2003; Carilli et
al. 2004; Riechers et al. 2007). 

Most recently, we have detected [CII] fine structure line emission
from J1148+5251 (Maiolino et al. 2005). Fine structure lines are the
dominant cooling mechanism for interstellar gas, and hence a key
diagnostic of the energetics of the interstellar medium. Figure 10
shows the high resolution image of the [CII] and CO emission from
J1148+5251.  The [CII] and CO emission are co-spatial, and extended
over a scale $\sim 5.5$ kpc, suggesting distributed gas heating, and
hence star formation, on kpc-scales.

\begin{figure}
\centerline{\psfig{file=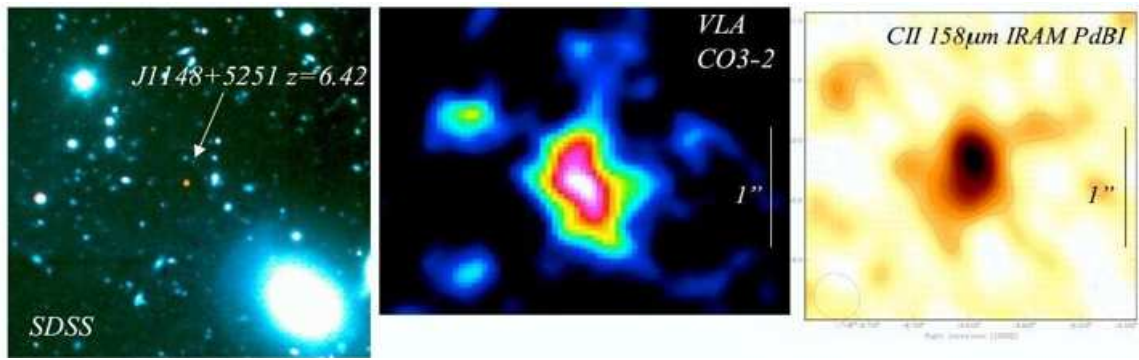,width=6in}}
\caption{Images of the most distant SDSS QSO J1148+5251 at z=6.42.
Left: The optical/nearIR SDSS color image (Fan et al. 2004).
Center: The CO 3-2 emission from the host galaxy observed
with NRAO's Very Large Array (Walter et al. 2004).
Right: The CII 158um emission observed with IRAM's Plateau de Bure
Interferometer   (Walter et al. in prep.).
This supermassive black hole (black hole mass ~ 1e9 solar masses)
is enveloped in a giant molecular gas cloud,
with a molecular gas mass ~ 2e10 solar masses, distributed over
a physical scale of about 5.5kpc. The extended CII emission
suggests active star formation across this region, with a total
star formation rate ~ 3000 solar masses per year (Maiolino et al.
2006). }
\end{figure}

Figure 11 shows the rest-frame radio through near-IR SED of J1148+5251
(Wang et al. 2007).  The source shows a clear excess in the FIR over
the expected SED of a typical low $z$, optically selected QSO. 
This SED from the FIR through the radio is consistent with that
expected for an active star forming galaxy with a dust temperature
of $\sim 50$K, and follows the radio-FIR correlation for 
star forming galaxies. Figure 4 also shows the CO excitation.
The gas excitation is reasonable fit by an LVG model implying
dense ($> 10^4$ cm$^{-3}$), warm ($\ge 50$K) molecular gas, 
comparable to what is seen in the centers of nearby nuclear
starburst galaxies (Bertoldi et al. 2004).

\begin{figure}
\psfig{file=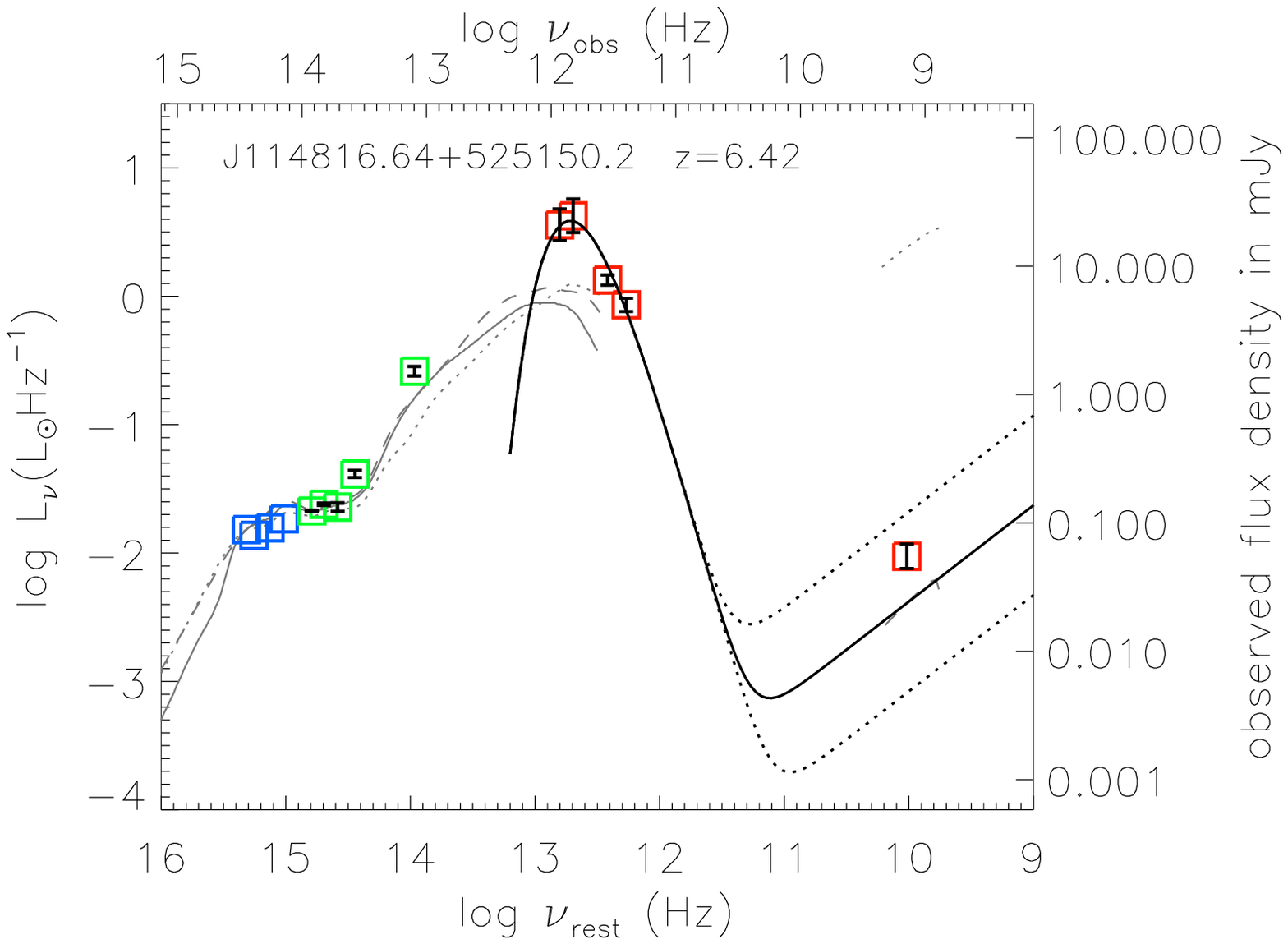,width=3.0in}
\vskip -2.0in
\hspace*{3.3in}
\psfig{file=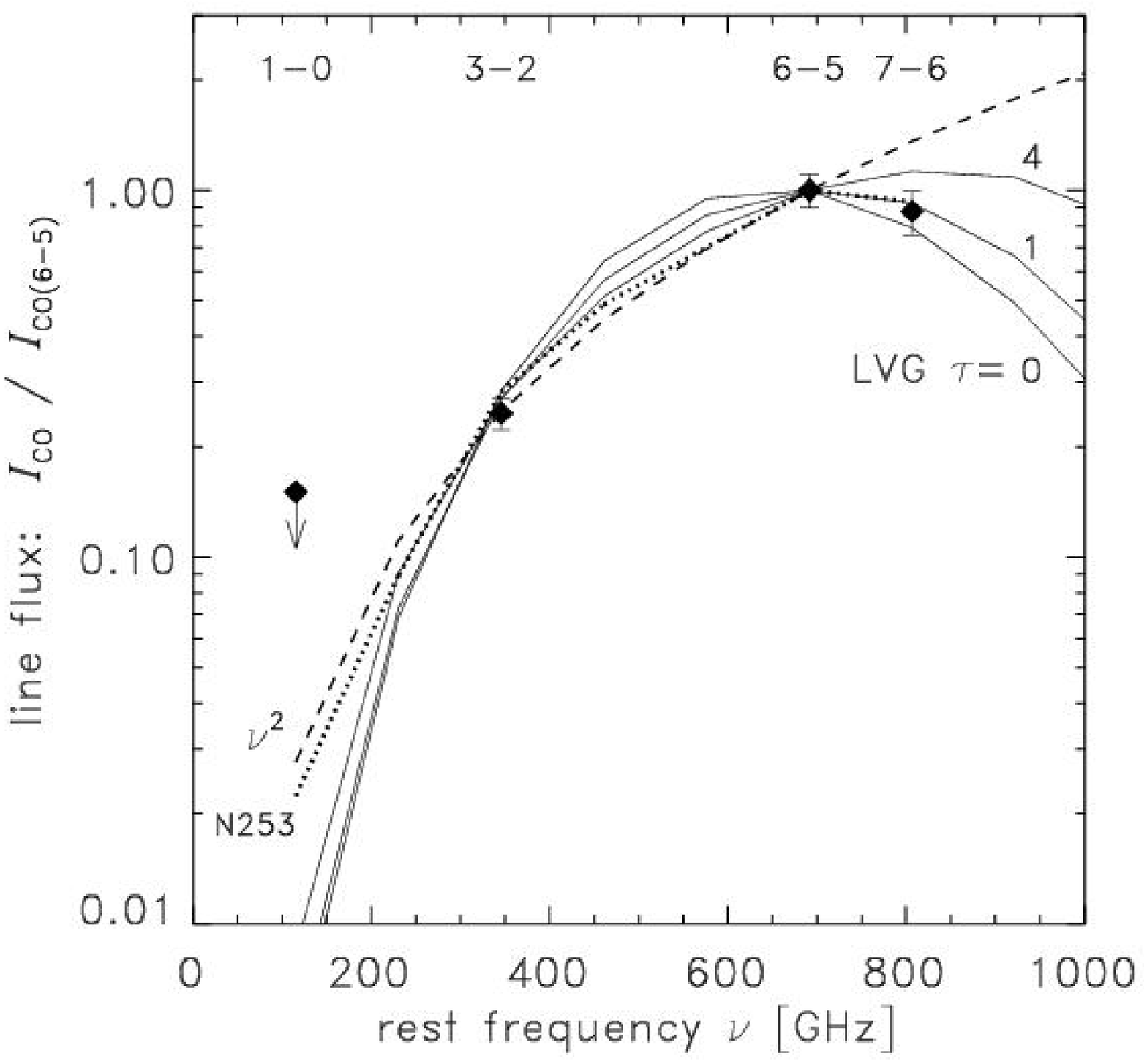,width=2.3in}
\caption{Left: The radio through near-IR SED of J1148+5251 (Wang
et al. 2007).
Right: The CO excitation in J1148+5215 (Bertoldi et al. 2003).}
\end{figure}

These observations of J1148+5251, and other $z \sim 6$ QSOs (Carilli
et al. 2007; Wang et al. 2007), demonstrate that large
reservoirs of dust and metal enriched atomic and molecular gas can
exist in massive galaxies within 1 Gyr of the Big Bang.  The current
observations suggest active star formation in the quasar host galaxy,
with a massive star formation rate of order $10^3$ M$_\odot$
year$^{-1}$, adequate to form a large elliptical galaxy in a dynamical
timescale $\sim 10^8$ years.

These observations of the host galaxies of the most distant quasars have
inspired a number of major theoretical efforts aimed at delineating
the formation of massive galaxies and supermassive blacks at the
earliest epochs. In particular, the theory groups at Harvard and
Arizona have performed state-of-the-art numerical simulations which
follow the formation of the largest galaxies in the early Universe. In
a series of recent papers (eg. Li et al. 2007; Narayanan et al. 2007),
they show that early massive galaxy and SMBH formation is possible in rare
peaks in the cosmic density field, through a series of gas-rich,
massive mergers starting at $z \sim 14$.  These systems evolve into
massive galaxies at the centers of the most massive cluster
environments seen today ($\sim 10^{15}$ M$_\odot$). These results are
generally consistent with the idea of 'downsizing' in both galaxy and
supermassive black hole formation (Cowie et al. 1996; Heckman et
al. 2004), meaning that the most massive black holes ($>10^9$
M$_\odot$) and galaxies ($> 10^{12}$ M$_\odot$) may form at high
redshift in extreme, gas rich mergers/accretion events, in rare peaks
in the cosmic density field.

\subsection{The Coming  Revolution: ALMA and a Phase I SKA}

While radio astronomy has pushed back into cosmic reionization through
these molecular line and continuum observations, current studies are
at the limit of the capabilities of existing millimeter and radio
telescopes. Even at these limits, these experiments stretch current
instrumentation to the extreme limit, such that only rare and
pathologic objects are detectable, ie. ultra- to hyper-luminous IR
galaxies, with star formation rates $\ge 100$ M$_\odot$ year$^{-1}$.
Fortunately, in the coming few years there will be dramatic
improvements in observational capabilities at submillimeter through
centimeter wavelengths that should revolutionize studies of the
earliest galaxies.

An order of magnitude, or more, increase in collecting area at mm and
cm wavelengths is required to enable the study of the first 'normal'
galaxies within the epoch of reionization (EoR), eg. the $z \ge 6$
Ly$\alpha$ galaxies currently being discovered in deep near-IR surveys
(SFRs $\sim 10$ to 100 M$_\odot$ year$^{-1}$; eg.
Murayama et al. 2007). At mm wavelengths this increase in capability
will be realized when the Atacama Large Millimeter Array becomes
operational early in the next decade. ALMA represents close to two
orders of magnitude improvement in all areas of submillimeter
astronomy, including: spatial resolution, spectral capabilities, and
sensitivity. At cm wavelengths, the Expanded Very Large Array opens up
the full frequency range from 1 to 50 GHz with order of magnitude
improved continuum sensitivity, and dramatically improved spectral
search capabilities. Unfortunately, the EVLA will still have the same
total collecting area as the VLA. An increase in collecting area by
roughly an order of magnitude over the EVLA is required at cm
wavelengths to match advances at other wavelength ranges, and
perform the key, complementary science afford at the short cm
wavelengths. 

The key areas of science explored by these new facilities 
includes: 

\begin{itemize}

\item {\it Dust and molecular gas:} ALMA, and a Phase I SKA (10\%
demonstrator $\sim 10\times$ EVLA collecting area) will reveal the
dust and molecular gas in samples of normal star forming galaxies at
$z > 6$, including Ly$alpha$ emitters and Ly-break galaxies.  These
observations constrain the timescale for metal and dust enrichment in
the most distant galaxies, and probe a key element in galaxy
formation, namely the fundamental fuel for star formation.

\item {\it Interstellar medium physics:} The physical conditions in
the ISM of these galaxies (density, temperature, abundances) can be
studied in detail, including gas excitation conditions, dust spectral
energy distributions, and observations of dense gas tracers such as
HCN and HCO+, directly associated with star forming regions.

\item {\it Fine structure lines:} The [CII] 158$\mu$m line, and other
atomic fine structure lines, are the dominant ISM gas cooling fine
structure lines.  These lines have great promise for exploring star
formation, ISM physics, and galaxy dynamics, in the first normal
galaxies.  I expect these lines to be the 'work-horse' lines for
studies of the first normal galaxies using ALMA. 

\item {\it Galaxy dynamics:} High resolution imaging of the
distribution of molecular and ionized gas is the most direct method to
determine the gas dynamics and dark matter content of the earliest
galaxies. These observations also present the only method with which
to test the M-$\sigma$ relation for galaxies and black holes out to
the highest redshifts.

\item {\it Star formation:} The centimeter continuum emission
associated with star formation in these galaxies can be observed,
including millarcsecond resolution imaging using Very Long Baseline
Interferometry.

\end{itemize}

\subsection{Continuum examples}

I illustrate the continuum studies with two recent examples. 
First, Figure 12 shows the results of a median stacking
analysis of a sample of 6500 $z \sim 3$ LBGs in the COSMOS field.  We
obtain a robust statistical detection at 1.4 GHz of $0.9 \pm 0.15$
$\mu$Jy.  Through the stacking analysis we are reaching sub-$\mu$Jy
sensitivity, ie. SKA-like depths, on galaxy populations at high
redshift. The interesting implication is that the median star
formation rate derived from the radio equals that derived from the
median rest-frame UV emission, assuming the standard factor five dust
correction for the UV emission. In other words, these data provide a
critical, independent check on the standard reddening factor for
high z LBGs (Carilli et al. in prep).  We have also identified the
first $z > 4$ submm galaxy, or massive dusty starburst, in these LBG
samples (Capak et al. in prep).

\begin{figure}
\psfig{file=LBG.U.ps,width=2.0in}
\vskip -2.2in
\hspace*{3.0in}
\psfig{file=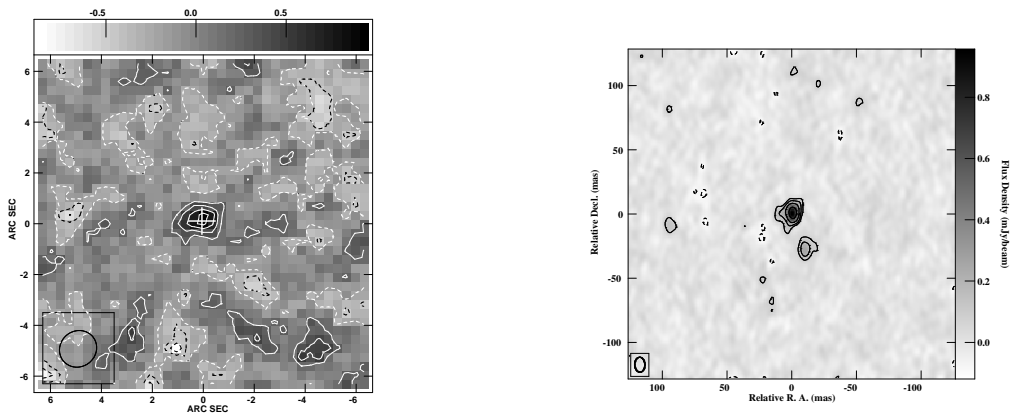,width=2.3in}
\hskip 0.2in
\caption{Left: A median stacked image at 1.4 GHz 
of 6500 Lyman Break galaxies from the COSMOS field 
($z \sim 3$, U drop-outs; Carilli et al. in prep). The contour levels are:
-0.75, -0.5, -0.25, 0.25, 0.5, 0.75 $\mu$Jy beam$^{-1}$.
Right: A VLBA image at 1.4 GHz resolution of the most 
distant radio loud AGN, J1427+3312 at $z = 6.12$. 
The contour levels are: -3,3,6,9,12 times the rms
noise of 28$\mu$Jy  beam$^{-1}$ (Momjian et al. in prep).
}
\end{figure}

Second, Figure 12 also shows a VLBA image with 28 $\mu$Jy sensitivity
of the most distant radio loud AGN, the $z = 6.12$ QSO J1427+3312
(McGreer et al. 2008; Momjian et al. 2008, in prep).  The source has a
steep spectral index of $\alpha_{1.4}^8 = -1.1$. The source is
comprised of two dominant components separated by $\sim 170$pc, with a
flux density ratio of about 4/1. Both components are clearly resolved
by these observations, with sizes $\sim 60$pc.  These data are
consistent with J1427+3312 being a Compact Symmetric Object. CSOs are
thought to be very young radio sources, $\le 10^5$ years, highly
confined by the dense ISM of the host galaxy (Taylor et al 1996;
Readhead et al. 1996). The conversion efficiency of jet kinetic energy
to radio luminosity is thought to increase in such highly confined
sources.
 
While it is dangerous to draw general conclusions from a single
source, we can speculate that at the highest redshifts a higher
fraction of radio jets will be CSOs, due to the higher gas densities
in the earliest galaxies. Kinetic feedback from such jets can affect
significantly the evolution of the host galaxy, perhaps inducing star
formation, and eventually driving enriched material into the IGM.
These radio source also present an ideal opportunity for observations
of HI 21cm absorption during cosmic reionization (Carilli et
al. 2002).

I should also point out that VLBA observations at 1.4 GHz are reaching
(rest frame) brightness temperature sensitivities at mas resolution
approaching 10$^5$ K (Momjian et al. 2007). This limit is critical,
since it is the dividing point between the brightness temperature
expected for an AGN versus that predicted for a starburst (Condon et
al. 1991).

\subsection{Complementarity and sensitivity}

There are a few important factors to keep in mind when considering 
SKA phase I studies of high $z$ molecular line emission, in comparison
to other, nearer term, telescopes:

\begin{itemize}

\item When considering the very first galaxies, into the EoR ($z >
6$), mm telescopes such as ALMA will be limited to studying very high
order transitions, eg. at $z=6$ the lowest ALMA band (80 -- 100 GHz)
redshifts to 690 GHz, corresponding to CO 6-5, or HCN 7-6.  It is
unclear that these higher order transitions will be excited in an
average galaxy, in particular for the high dipole moment molecules
such as HCN, since the required densities for excitation become
extreme ($>> 10^5$ cm$^{-3}$).\footnote{The CMB will also contribute
to excitation.  For example, at z=6, T(CMB) = 19K, while the typical
CO excitation temperature in an active star forming galaxy is $\sim
30$ to 40K.}  For comparison, in normal galaxies, such as the Milky
Way, constant T$_B$ holds for the integrated CO emission roughly up to
CO 3-2, above which the lines are sub-thermally excited.

\item The SKA phase I design corresponds to a collecting area roughly
an order of magnitude larger than the EVLA. While the EVLA continuum
sensitivity at higher frequencies is a factor 9 improvement over the
existing VLA due to the increased bandwidth, the EVLA will have the
same collecting area as the VLA, and hence will have at best
marginally improved sensitivity for spectral line (ie. band-limited)
studies, perhaps 50\% due to small improvements in receivers and
antenna optics. What the EVLA does provide for line science is
improved search capabilities, in terms of new receiver bands and much
wider bandwidths with many more channels.

\item It should be emphasized that the natural high frequency
limit for cm telescopes is $\sim 45$ GHz, as set by the atmospheric
O$_2$ line. If the SKA phase I was extended to this natural limit, the
effective sensitivity is increased by another factor four for thermal
objects (assuming constant brightness temperature) relative to a 22
GHz design, improving the SKA phase I to a 40\% SKA.  A higher
frequency would also decrease the necessary longest baseline by a
factor two at fixed resolution, thereby reducing the costs of array
configuration, connectivity, and data transmission.

\item Beyond pointed observations of known
high-$z$ objects, the much wider field of view and fractional
bandwidth of the SKA relative to eg. ALMA, will allow for efficient
surveys for molecular emission line galaxies from $z \sim 2$ into the
reionization (Carilli \& Blain 2003).

\end{itemize}

The key point is that, for the first galaxies, the fine structure
lines, such as [CII] 158$\mu$m will be the work-horse lines for ALMA,
while studies of molecular gas, the fundamental fuel for star
formation in galaxies, will be the regime of short cm wavelength
telescopes.  

Figure 13 shows the sensitivity to molecular line emission for the
EVLA, the SKA phase I, and ALMA.  Also shown is a model for the
expected continuum and CO line emission from an active star forming
galaxy with a total IR luminosity $\sim 10\%$ that of Arp 220 at $z =
5$, corresponding to a star formation rate of a few 10's M$_\odot$
 year$^{-1}$. Such a galaxy would be comparable to the Ly$\alpha$
galaxies seen in deep optical surveys at $z \sim 6$,
ie. characteristic of the high $z$ galaxy population (Murayama et al.
2007). I have assumed a CO excitation ladder similar to that seen in
the few high $z$ molecular line emitting galaxies currently known
(this may be biased to higher excitation).

\begin{figure}
\centerline{\psfig{figure=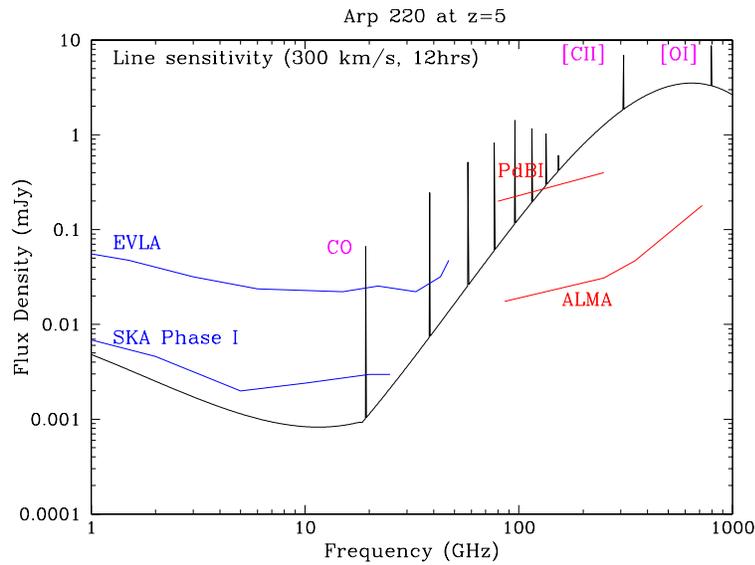,width=4in,angle=-90}}
\caption{The dotted curves show the (1$\sigma$) sensitivity to
spectral line emission for the VLA, SKA phase I, and ALMA in 12
hours. The solid curve is the emission spectrum, including continuum
and CO lines, for a source with the luminosity of Arp 220
(ie. a source with FIR $\sim 10^{12}$ L$_\odot$) at $z = 5$. }
\end{figure}

Figure 1 shows that the SKA phase I will be able to detect the CO 1-0
and 2-1 emission from  'normal' star forming galaxies out to
very high redshifts, into the epoch of cosmic reionization.  ALMA will
easily detect the higher order transitions, in sources where they are
excited. ALMA will also be able to study the low excitation fine
structure line emission from primeval galaxies ([CII] 158$\mu$m, [OI]
63$\mu$m).

For completeness, in Figure 14, I show the continuum sensitivities of
these telescopes. The high frequency SKA phase I will easily detect
the continuum emission over the rest frame frequency range of 50 to
200 GHz range. The lower end of this range is where free-free emission
will dominate, providing the most direct, dust-unbiased, measure of
the star formation rate.  And the upper end corresponds to the cold
dust emission.
 
\begin{figure}
\centerline{\psfig{figure=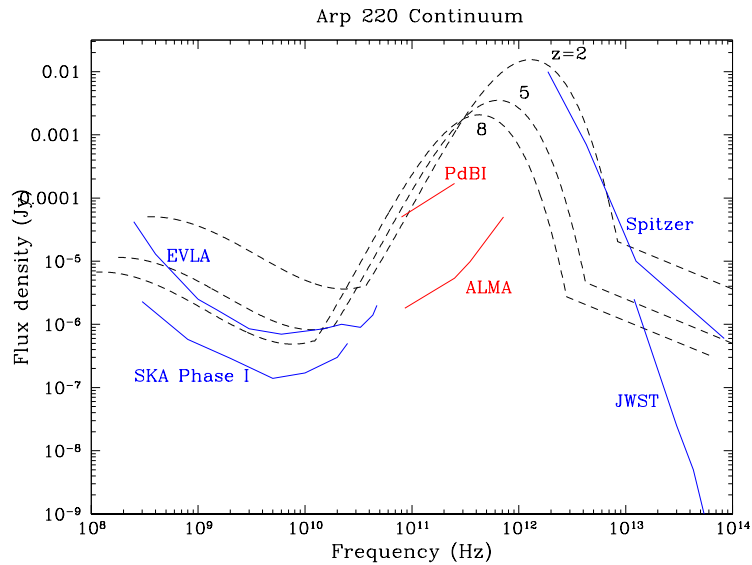,width=4in,angle=-90}}
\caption{The solid curves show the (1$\sigma$) sensitivity to continuum
emission for the EVLA, SKA phase I, ALMA, Spitzer, and the JWST,
in 12 hours. The dashed curves are the continuum emission 
spectrum for a source with the luminosity of Arp 220
(ie. a source with FIR $\sim 10^{12}$ L$_\odot$) at
$z = 2$, 5, and 8. 
}
\end{figure}

Overall, the EVLA Phase III and ALMA provide the necessary
complement to large ground and space-based optical and IR
telescopes, enabling a panchromatic study of the formation of normal
galaxies back to the EoR.  

\begin{acknowledgments}
CC thanks the Max-Planck-Gesellschaft and the Humboldt-Stiftung for
support through the Max-Planck-Forschungspreis.  The National Radio
Astronomy Observatory is a facility of the National Science
Foundation, operated by Associated Universities, Inc..  Thanks also to
N. Gnedin, M. McQuinn, M. Zaldariagga, S. Wyithe, W. Lane, W. Cotton,
A. Chippendale, E. Momjian, F. Walter for permission to reproduce figures.
\end{acknowledgments}

{}

\end{document}